\documentclass[cameraready]{Interspeech}
\usepackage{listings}
\usepackage{graphicx}
\usepackage{subcaption}
\usepackage{listings}
\title{Benchmarking Speech Systems for Frontline Health Conversations: \\ The DISPLACE-M Challenge\thanks{This work was funded by Ministry of Electronics and Information Technology (MeitY),
Government of India through the National Language Translation Mission
(NLTM): BHASHINI project, SP/MITO-22-001 grant. The data collection was funded by Ministry of Education through AI Center of Excellence in Health grants.}
\thanks{Submitted for review to Interspeech 2026.}}

\author[affiliation={1}]{Dhanya}{E}
\author[affiliation={1}]{Ankita}{Meena}
\author[affiliation={1}]{Manas}{Nanivadekar}
\author[affiliation={1}]{Noumida}{A}
\author[affiliation={1}]{Victor}{Azad}
\author[affiliation={2}]{Ashwini Nagaraj}{Shenoy}
\author[affiliation={2}]{Pratik}{Roy Chowdhuri}
\author[affiliation={3}]{Shobhit}{Banga}
\author[affiliation={3}]{Vanshika}{Chhabra}
\author[affiliation={4}]{Chitralekha}{Bhat}
\author[affiliation={5}]{Shareef babu}{Kalluri}
\author[affiliation={6}]{Srikanth Raj}{Chetupalli}
\author[affiliation={2}]{Deepu}{Vijayasenan}
\author[affiliation={1,4}]{Sriram}{Ganapathy}
\address{
    $^1$ LEAP lab, EE, Indian Institute of Science, Bangalore, India \\
    $^2$ National Institute of Technology Karnataka, India \\
    $^3$ Josh Talks, India \\
    $^4$ TANUH AI Center of Excellence, Bangalore, India \\
    $^5$ UPES, India \\
    $^6$ Indian Institute of Technology Bombay, India 
}

\email{dhanyae@iisc.ac.in}


\usepackage{comment}

\begin{document}

\maketitle

\begin{abstract}
    %1188 characters
The DIarization and Speech Processing for LAnguage understanding in Conversational Environments - Medical (DISPLACE-M) challenge introduces a conversational AI benchmark for understanding goal-oriented, real-world medical dialogues. The challenge addresses multi-speaker interactions between frontline health workers and care seekers, characterized by spontaneous, noisy and overlapping speech. As part of the challenge, medical conversational dataset comprising $40$ hours of development and $15$ hours of blind evaluation recordings was released. We provided baseline systems across 4 tasks - speaker diarization, automatic speech recognition, topic identification and dialogue summarization - to enable consistent benchmarking. System performance is evaluated using diarization error rate (DER), time-constrained minimum-permutation word error rate (tcpWER) and ROUGE-L. This paper describes the Phase-I evaluation - data, tasks and baseline systems - along with the summary of the evaluation results.

%During this evaluation (Phase-I), 
% In Phase-I evaluation, several teams, across the globe, actively participated pushing the baseline results.
% and the task is shown to be substantially challenging, and the existing systems are significantly short of healthcare deployment readiness.
%systems on these metrics. However, even with a $6-8$ week dedicated effort from various participants, the task is shown to be substantially challenging, and the existing systems are significantly short of healthcare deployment readiness.
\end{abstract}

\section{Introduction}
\label{sec:into}

Speech processing tools for the understanding of conversations from the healthcare domain can have transformative effects on public health systems, especially in community healthcare settings. Existing speech datasets in this domain are mostly recorded in controlled environments such as hospitals, involve structured clinician-patient conversations, and are predominantly in English. AI-driven tools developed using such datasets often fail to perform when deployed in real-world setups involving community healthcare workers. Further, there is a lack of research in multi-lingual settings. To catalyze research efforts in this space, we collected and released an annotated corpus of conversations between community health workers and healthcare seekers in informal settings in Hindi. Using the data, we hosted an open leaderboard-style challenge - ``The \textbf{DI}arization and \textbf{S}peech \textbf{P}rocessing for \textbf{LA}nguage understanding in \textbf{C}onversational \textbf{E}nvironments - Medical (DISPLACE-M) challenge''. 
This paper makes three main contributions:
\begin{enumerate}
    \item A new benchmark built from Hindi health conversations capturing spontaneous, code-mixed, multi-speaker interactions recorded in unconstrained natural settings.
    \item A unified evaluation framework covering four interconnected tasks - (i) speaker diarization, (ii) automatic speech recognition (ASR), (iii) topic identification, and (iv) dialogue summarization, to perform end-to-end assessment of conversational speech-understanding systems. 
    \item Baseline systems and evaluation frameworks to support reproducible research and a leaderboard platform to propel subsequent phases of the challenge.
    % \textit{We also provided the leader-board platform for all the four tracks for the continuous monitoring of the participants progress in system development. }
\end{enumerate} 

%\begin{figure*}
%	\centering
%	\includegraphics[width=\linewidth]{figures/DISPLACE-M_timeline.png}
%	\caption{Timeline of DISPLACE-M}
%	\label{fig:timeline}
%\end{figure*}

\section{Related Work and Paper Organization}
\label{sec:related_work}

ASR research has seen significant progress in conversational settings, particularly for meetings, call centers, and other multi-party environments. Datasets such as the ICSI Meeting Corpus \cite{ICSI_Meeting_dataset} and the AMI Meeting Corpus \cite{ami_datasset} have been widely used to study spontaneous multi-speaker conversational speech. 
% The ICSI corpus consists of technical discussion recorded in controlled indoor settings. The AMI Meeting Corpus contains multi-party meeting recordings with rich annotations, including transcripts, speaker labels, and dialogue acts and has been widely used for research on meeting transcription, topic segmentation, and conversational analysis.  
However, the conversations in these datasets differ significantly from those in frontline healthcare settings, where conversations are goal-driven, medically grounded, and often involve nuanced contextual information.  Topic classification has been widely studied for meetings, call centers, and online discussions, while dialogue summarization datasets have largely been derived from customer support conversations, news interviews,  scripted meeting scenarios or informal meetings  \cite{gliwa-etal-2019-samsum, Zhong2021QMSumAN, baghel2024summary, kalluri2024second}. 

Research in spoken conversation processing in healthcare has primarily focused on data collected in clinical or hospital environments. Resources such as MIMIC \cite{Johnson2023} and i2b2 \cite{i2b2_nlp_datasets} 
%(Informatics for Integrating Biology and the Bedside) 
have enabled extensive work on clinical documentation and information extraction, but these datasets are predominantly English and reflect formal clinical workflows rather than informal frontline interactions \cite{Uzuner2015}. Similarly, prior ASR efforts have focused on physician dictation or structured clinical interviews recorded under relatively clean acoustic conditions with limited speaker variability \cite{chiu18_interspeech}. These works largely focus on ASR evaluations and do not   probe goal-oriented conversational metrics. 

In the Indian context, there has been a growing interest in developing diarization and ASR systems for multilingual settings~\cite{baghel2023displace}. However, existing datasets are restricted to isolated utterances or short conversational clips \cite{Bhogale2022EffectivenessOM, bhogale23_interspeech, Nithya2023SPRINGINXAM, javed-etal-2024-indicvoices}, which limits their suitability for studying long-form, dialogue centric ASR systems. Further, there are no public datasets covering conversations in the healthcare domain.
% generally evaluated on read speech, broadcast media or short conversational clips \cite{Bhogale2022EffectivenessOM, bhogale23_interspeech, Nithya2023SPRINGINXAM, javed-etal-2024-indicvoices} rather than long, task-oriented health care dialogues. For example, IndicVoices \cite{javed-etal-2024-indicvoices} includes read, extempore, and conversational speech, but the conversations are relatively brief, typically 2-3 minutes long, which limits their suitability for studying long-form, dialogue centric ASR systems. 
% Beyond transcription, higher-level understanding tasks such as topic identification and dialogue summarization have received considerable attention in other domains. 
Despite substantial progress in medical NLP and conversational ASR, there is still no benchmark that captures the complexity of real-world frontline healthcare interactions. 
% Moreover, prior datasets usually focus on a single task, such as transcription or summarization, rather than supporting end-to-end conversational understanding. 
% The DISPLACE-M Challenge addresses these limitations by introducing a multi-task benchmark spanning diarization, ASR, topic identification, and dialogue summarization.
\begin{figure*}[t]
  \centering
  \includegraphics[width=0.95\textwidth]{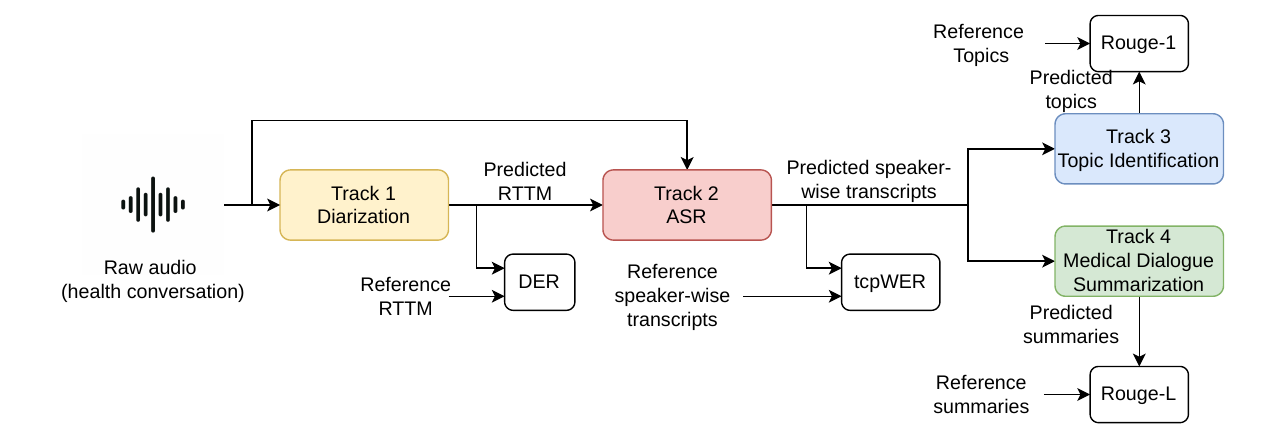}
  \vspace{-0.5em} 
  \caption{DISPLACE-M Baseline system development pipeline which involves a cascaded approach of speaker diarization (SD), ASR and topic identification / summarization modules.}
  \label{fig:baseline_pipeline}
  \vspace{-0.05in}
\end{figure*}

To fill this gap, we collected the DISPLACE-M dataset and launched the evaluation campaign. The Phase-I of the DISPLACE-M evaluations was organized between January 1, 2026 and February 27, 2026. %Timeline is shown in Figure~\ref{fig:timeline}. 
The details of the  dataset (Section~\ref{sec:dataset}), tracks and metrics (Section~\ref{sec:tracks}), baseline systems (Section~\ref{sec:disp_baseline}) and challenge results (Section~\ref{sec:challenge_results}) are described below.

\section{DISPLACE-M Challenge}
\label{sec:dataset}
\subsection{Data Collection}
\label{ssec:dataset_details}
% The medical conversational corpus has been collected in collaboration with an external vendor, who is responsible for on-ground data acquisition. The goal of this data collection exercise is to build a diverse, high-quality dataset to support the development and evaluation of AI models that can assist in healthcare tasks such as information extraction, summarization, and report generation. 
% The primary goal of our data collection exercise was to build a diverse, high-quality, medical domain-specific, conversational speech dataset to support the development and evaluation of speech analytics solutions, such as automatic summarization, report generation, transcription, and information extraction.

We recorded around $55$ hours of annotated healthcare conversational speech from rural and semi-urban areas, collected through field deployments in the Indian
states of Haryana and Bihar. Data collection was conducted by $80$ frontline community health workers across $20$ administrative blocks spanning $10$ districts. The primary participants were NPHWs (non-physician health workers) -  ASHA\footnote{ASHA (Accredited Social Health Activists) workers, who are government-appointed female community health workers serving as the first point of contact for healthcare seekers (HS)} and Anganwadi Sevikas \footnote{Anganwadi sevikas are frontline workers under India’s Integrated Child Development Services (ICDS) program} and healthcare seekers (HS). Recordings captured natural NPHW-HS conversations occurring during routine community healthcare interactions. 
These interactions took place in real care environments such as Primary Health Centres (PHCs), Anganwadi centers, primary schools, open village meeting spaces (chaupals), household visits and community outreach settings. In our recordings, Hindi served as the primary language with code-switching to Indian English and with heavy use of regional dialects, such as Haryanvi, Bhojpuri, and Magahi, reflecting natural linguistic practices.

The recorded conversations include speakers from different age groups. Each recording ranges from $5$-$30$ minutes, typically involving goal-oriented conversations between two participants - NPHW and a healthcare seeker. The conversations, which include nature field ambiance, 
primarily focused on topics such as General Health and Constitutional Symptoms (e.g., weakness, fatigue, weight loss, low hemoglobin/blood loss), Women’s Health and Gynecological Issues (e.g., menstrual problems, pregnancy, family planning), Acute Illnesses and Infections (e.g., fever, cough, cold, typhoid, dengue), and Preventive Care and Diagnostics (e.g., routine check-ups, blood tests, ultrasounds).
All audio recordings were captured using the far-field microphone of a mobile device. 
The recording format was WAV using $16$-bit PCM at $16$kHz sampling rate. 
% Background sounds are limited to natural community noises (for example, household, clinic, or outdoor environments) that do not obscure intelligibility for a human listener.

\subsection{Annotation}
\label{ssec:data_annotation}
All recordings were processed through a structured, multi-stage manual annotation pipeline prior to dataset finalization. We performed a multi-stage check to ensure quality of the audio files, accuracy of the time-markers and the correctness of the ASR transcripts. Further, the clinical summary is given by an expert clinician (Doctor) who had access to the audio and ground-truth rich transcriptions. 
The workflow was organized into distinct segmentation, transcription, and quality assurance phases to ensure traceability and reproducibility. 
More details of the dataset, pre-processing steps, and the annotation quality-checks are provided in the supplementary material. 

\subsection{Development and Evaluation set}
\label{ssec:disp_dev_eval}
The development and evaluation set contains $35$ hours of Hindi medical conversations involving $260$ unique speakers, comprising both HS and NPHWs. The speakers span a broad age range from $19$-$80$ years. %Individual recordings vary in duration from approximately $5$-$20$ minutes \textcolor{red}{In the above description we are saying 15-30mts}.

For Tracks 1 (speaker diarization) and 2 (ASR), the dataset is split into $25$ hours for development and $10$ hours for blind evaluation. The development data was released in two phases: Dev Set 1 ($15$ hours) and Dev Set 2 ($10$ hours). For Tracks 3 (topic identification) and Track 4 (dialogue summarization), a separate curated subset of about $15$ hours is provided, with $10$ hours for development and $5$~hours for evaluation. 

Each track has a separate set of reference annotations. 
The output hypotheses generated by the challenge participants were instructed to follow a specific format (like RTTM, json) for each task in order to enable the evaluation in the open-leaderboard\footnote{\url{https://www.codabench.org/competitions/13833/?secret\_key=1b714e64-0f0d-4e0f-8a3c-be9b3d10f00c\#}}.

% Diarization labels are provided in RTTM format. 
% For the ASR track, speech segment files are provided indicating speech regions and speaker segments, along with corresponding transcripts in Hindi. For the downstream language understanding tracks, ground-truth topic labels are provided for Track 3 and reference medical summaries prepared by trained medical practitioners are provided for Track 4.

\section{Challenge Tracks and Metrics}\label{sec:tracks}
% The DISPLACE-M dataset captures frontline healthcare conversations, recorded using mobile phone-like devices. Unlike studio-quality or meeting-style recordings, the recordings contain background noise, speech overlaps, and natural turn-taking. The challenge provides an open, standardized testbed for evaluating the robustness of diarization and ASR systems in multi-accent, multi-dialect, and noisy conversational conditions, a critical gap in current evaluations. 

We define four interconnected tasks -: 
\begin{enumerate}
\item \textbf{Track 1 - Speaker Diarization (SD)} aims to determine ``who spoke when,'' by automatically segmenting the audio into speaker-homogeneous regions. The performance metric is the diarization error rate (DER). 
\item \textbf{Track 2 - Automatic Speech Recognition (ASR)} system transcribes multi-speaker   healthcare conversations. Systems are expected to generate time-marked word-level transcriptions for the spontaneous conversational speech. The performance metric is character error rate (CER), word-error rate (WER) and tcpWER (Time-Constrained minimum-Permutation Word Error Rate)~\cite{neumann23_chime}. The tcpWER extends standard WER to multi-speaker settings by computing the minimum WER over all possible permutations of reference and hypothesized speaker streams after concatenating each speaker’s utterances \cite{neumann23_chime} with constraints along the temporal axis for alignment. It therefore jointly reflects transcription quality and speaker assignment consistency, making it suitable for evaluating end-to-end diarization-aware ASR.

\item \textbf{Track 3 - Topic Identification (TI)} aims to identify the underlying topics discussed between the NPHW and the HS. This is similar to topic modeling in NLP applications. The performance metrics are Rouge-1 (R-1) and Rouge-L (R-L)~\cite{lin-2004-rouge}. Specifically, R-1 measures the overlap of individual clinical keywords (unigrams), while R-L identifies the longest common subsequence to account for sentence structure and word order.
% Given an audio, recording  systems are required to classify or tag one or more predefined topics relevant to the conversation. This track evaluates a system’s ability to capture semantic and contextual information from noisy, real-world conversational data.
\item \textbf{Track 4 -  Dialogue Summarization (DS)} focuses on generating concise and informative summaries of multi-speaker conversations, preserving the essential medical  context.  The performance metric is the Rouge-L score~\cite{lin-2004-rouge}. 
\end{enumerate}

\section{Baseline Systems}
\label{sec:disp_baseline}

% We implemented baseline systems for all four tracks of the DISPLACE-M Challenge. 
For the baseline system development, we follow a cascaded system approach illustrated in Figure~\ref{fig:baseline_pipeline}. Given raw conversational audio, we first perform speaker diarization to obtain time-stamped speaker segments. These segments are subsequently used to guide the ASR module, which generates speaker-attributed transcripts. These %resulting 
transcripts then serve as inputs to  topic identification and dialogue summarization.

%----------------------- SPEAKER DIARIZATION ---------------------------
\subsection{Track-1: Speaker Diarization (SD)}
The speaker diarization system is the base variant of the DiariZen\footnote{https://github.com/BUTSpeechFIT/DiariZen} model \cite{han2025leveraging,han2025fine,han2025efficient}, which
%developed by Speech@FIT at Brno University of Technology. 
%Diarizen\footnote{https://github.com/BUTSpeechFIT/DiariZen} 
closely follows the pyannote \cite{bredin2023pyannote} pipeline. The pipeline follows a two-stage process, in which an end-to-end neural diarization (EEND) model first computes the local diarization results on segments of $8$s duration, followed by an Agglomerative Hierarchical Clustering (AHC) stage, where the segment-wise speaker embeddings are grouped to produce the final diarization output. DiariZen uses a custom architecture for the EEND model trained using the powerset loss \cite{plaquet2023powerset}, and 
% with the local end-to-end neural diarization (EEND) model replaced by a custom architecture trained using the powerset loss \cite{plaquet2023powerset}. 
% \textit{The model breaks the input audio into 8s chunks of overlapping segments, and passes it to the EEND module to obtain the local diariztion results. These results are combined by finding a mapping between the speakers in the segment-wise diariztion ouputs. This is achieved by extracting speaker embeddings for each speaker in the segments and clustering them using Agglomerative Hierarchical Clustering (AHC). 
ResNet34-LM trained on the VoxCeleb2 dataset \cite{chung2018voxceleb2} to extract the speaker embeddings. AHC is carried out with  the number of clusters set to $2$, the minimum cluster size set to $13$, and an AHC threshold of $0.6$. This system is referred to as Baseline-1.% this can be removed 

For the DISPLACE-M baseline, DiariZen was evaluated under two setups - zero-shot inference (Baseline-1) and supervised fine-tuning with the development data (Baseline-2). The fine-tuning was done %on both DEV1 and DEV2 using 5-fold cross-validation, and 
with the AHC threshold set to $0.7$ and minimum cluster size set to $30$. %This system is referred to as Baseline-2.
\subsection{Track-2: Automatic Speech Recognition (ASR)}
\label{ssec:asr-baseline}
We released two baseline systems based on i) IndicConformer\footnote{https://github.com/AI4Bharat/IndicConformerASR} and ii) Whisper-large-v3\footnote{https://huggingface.co/openai/whisper-large-v3}. 
\begin{enumerate}
    \item IndicConformer: This model was developed by the AI4Bharat initiative. We used \textit{indic-conformer-600m-multilingual} model, a Conformer-based end-to-end ASR model trained on large-scale multilingual Indian speech corpora covering 22 Indian languages \cite{javed-etal-2024-indicvoices}
    %. We used the \textit{indic-conformer-600m-multilingual} model 
    for the first baseline.  
    \item Whisper-large-v3: The second baseline uses Whisper-large-v3 \cite{radford2023robust}, a 1550-M parameter multilingual encoder–decoder speech foundation model trained on web-scale weakly supervised audio.
\end{enumerate}
%For the DISPLACE baseline, both models were evaluated under two settings: zero-shot inference and supervised fine-tuning on the challenge development set. 
Similar to Track-1, both the models were evaluated with and without fine-tuning on the development set. 
% We note that, fine-tuning led to substantial improvements in recognizing colloquial Hindi, code-mixed terminology and domain-specific medical expressions.

% \textbf{Evaluation Metric:} Systems were evaluated using Character Error Rate(CER) and Word Error Rate (WER) computed on normalized (removal of punctuation, standardization of numerals) Hindi transcripts. In addition, we report Time-Constrained minimum-Permutation Word Error Rate (tcpWER)\footnote{https://github.com/fgnt/meeteval} to jointly capture transcription accuracy and speaker attribution errors arising from diarization. tcpWER extends standard WER to multi-speaker settings by computing the minimum WER over all possible permutations of reference and hypothesized speaker streams after concatenating each speaker’s utterances \cite{neumann23_chime} with constraints along the temporal axis for alignment. It therefore jointly reflects transcription quality and speaker assignment consistency, making it suitable for evaluating end-to-end diarization-aware ASR systems.

\subsection{Track-3: Topic Identification (TI)}
\label{ssec:ti-baseline}
We employ an ASR–LLM cascade pipeline to identify medical topics in healthcare conversations. This approach first converts spontaneous speech into text using ASR systems, followed by an LLM - \texttt{medgemma-1.5-4b-it}~\cite{sellergren2025medgemma} - 
to generate the relevant health topics.
The specific prompt and setting used is described in Appendix~\ref{ssec:task3_topic_id} %the supplementary material. 

% We chose an LLM rather than a traditional classifier, since LLMs support open-set topic identification, identifying previously unseen topics through zero-shot reasoning. This is particularly advantageous in rural settings where dialogues are often informal and code-mixed. The separation of speech recognition and semantic analysis provides a flexible framework that benefits from independent advancements in ASR technology.

% \textbf{Evaluation Metrics}:
% Topic identification performance is assessed by comparing LLM predictions against expert annotations using a dual-metric strategy. To quantify lexical overlap, we report ROUGE-1 (R-1) and ROUGE-L (R-L) \cite{lin-2004-rouge}. Specifically, R-1 measures the overlap of individual clinical keywords (unigrams), while R-L identifies the longest common subsequence to account for sentence structure and word order. However, since zero-shot models often generate semantically correct synonyms rather than exact word-for-word matches, we also employ BERTScore \cite{Zhang*2020BERTScore}. This metric utilizes contextual embeddings to robustly evaluate semantic similarity.

\subsection{Track 4: Dialogue Summarization (DS)}
\label{ssec:ds-baseline}
Similar to the topic identification pipeline, we used the ASR transcript for the dialogue summarization (DS) task. 
% We employed a cascaded architecture based on ASR baseline, which serves as a strong baseline. In this cascaded architecture, speech is first transcribed into text using an ASR system, and the resulting transcript is then summarized using a text-based summarization model. 
For the summarization stage, we used \texttt{LLAMA-3.2-3B} model~\cite{mansha2025resource} as the text-based summarization model in a zero-shot setting. The prompt used for summarization is described in  Appendix~\ref{ssec:task4_ds} %the supplementary material.

% This design leverages the strong performance of modern ASR systems and large language models (LLMs), while keeping speech recognition and summarization as two clearly separated stages. 
% The cascaded ASR-based approach remains effective in practice due to the strong performance
% of modern ASR systems and large language models used for text summarization.

% \textbf{Evaluation Metric:} To evaluate the quality of generated summaries, we measure both lexical overlap and semantic similarity with reference summaries. For lexical evaluation, we use the standard ROUGE metrics \cite{lin-2004-rouge}, which quantify the overlap of
% $n$-grams or subsequences between the generated summary and the reference summary. Specifically, we report ROUGE-L (R-L). To capture semantic similarity beyond exact word overlap, we also use BERTScore \cite{Zhang*2020BERTScore}. which computes the similarity between contextual embeddings of the generated and reference summaries.

\section{Challenge Results}
\label{sec:challenge_results}
The DISPLACE-M Challenge received registrations from multiple teams across academia and industry, demonstrating a broad interest in the proposed benchmark. $12$ international teams actively participated in the Phase-I leaderboard  evaluations. 

We also benchmarked state-of-the-art closed-source models as reference systems. In particular, we evaluated Gemini 2.5 Pro \cite{comanici2025gemini} across all four tracks and Sarvam AI’s Saaras v3\footnote{\url{https://dashboard.sarvam.ai/speech-to-text}} for Track 1 (Speaker Diarization) and Track 2 (ASR). %(Automatic Speech Recognition).
 
\subsection{Track 1: Speaker Diarization}
\begin{figure}[t!]
  \centering
  \includegraphics[width=\columnwidth]{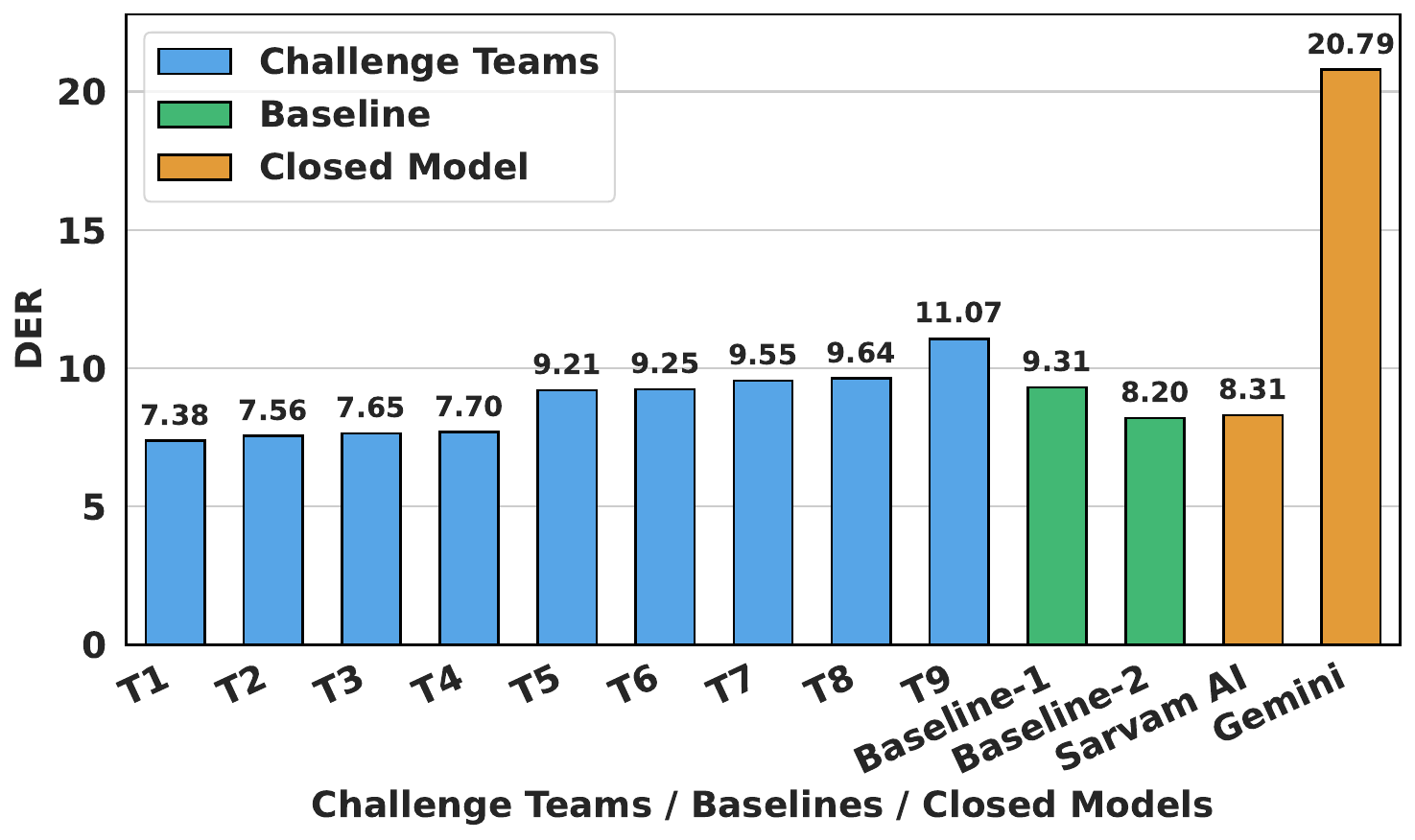}
  \caption{Comparison of Diarization Error Rate (DER \%) for Track 1 Evaluation Phase-I DS task.}
  \label{fig:track1_sd_res}
  \vspace{-0.1in} 
\end{figure}

Figure~\ref{fig:track1_sd_res} illustrates the Eval Phase-I results for the SD track. A total of $9$ teams participated in Phase-I evaluation. The top four systems achieved  improvements over Baseline-2,.

Team 1 (T1) used a hybrid end-to-end system trained on multiple datasets and fine-tuned it on the development sets. They also employed a Dynamic Logits Fusion strategy to combine 5 complementary systems, and obtained their best results.
% The top-five systems obtained a DER of 7.38\%, 7.56\% and 7.65\%, respectively. 
Team 2 (T2) and  Team 3 (T3) fine-tuned the large variant of DiariZen with the WavLM model. T2 used Dev-1 for training and Dev-2 for validation, and obtained their best results with learning rates of $10^{-3}$ and $2 \times 10^{-5}$ for the conformer blocks and WavLM, respectively. In contrast, T3 used a mix of both the development sets, employing a $80:20$ train-validation split. They utilize a three-stage progressive fine-tuning strategy with different hyperparameter values in each stage. 

% Team 4 modified the base Streaming Sortformer (SSF) model to cater to 2 two speakers, and fine-tuned it using randomly sampled 90s segments from both the development sets.
% % Team 4 fine-tuned the Streaming Sortformer (SSF) on both the development sets, with the output layer modified to two speakers. 
% Additionally, to improve boundary detection accuracy, they add a single 1-D convolution layer on top of the encoder to alter the frame resolution of SSF to 10ms.

\subsection{Track 2: ASR}
The ASR track %(Track-2) 
received $4$ submissions, of which two teams outperformed the baseline system. Table \ref{tab:asr_results} presents the Eval Phase-I results. %for the ASR track. 
Among the baselines described in Section \ref{ssec:asr-baseline}, fine-tuning significantly improved performance over the zero-shot setting. The fine-tuned IndicConformer achieved a tcpWER of $20.23$\% compared to $26.78$\% in the zero-shot setup, indicating the importance of in-domain adaptation. 

Team 1 (T1) fine-tuned the Qwen3-ASR-1.7B model \cite{Qwen3-ASR} using approximately 1,800 hours of open-source Hindi speech data, along with the challenge development set, to better adapt to domain-specific conversational speech. An LLM-based post-processing module using GPT-4.1 was further employed for error correction, with a particular focus on improving the recognition of medical terminology. Team 2 (T2) followed an approach similar to our Baseline-2 system, fine-tuning IndicConformer model on the challenge development set.

\begin{table}[t]
\centering
\caption{Results on Track-2 (ASR); ZS = Zero-shot inference, FT = Fine-tuned on challenge dev-set.}
% \vspace{-1em} 
\label{tab:asr_results}
\resizebox{\columnwidth}{!}{
\begin{tabular}{lccc}
\hline
\textbf{System} & \textbf{CER (\%)} & \textbf{WER (\%)} & \textbf{tcpWER(\%)} \\
\hline
\multicolumn{4}{l}{\textit{Challenge Submissions}} \\
T1 & \textbf{10.59} & \textbf{18.15} & \textbf{18.63} \\
T2 & 11.92 & 19.26 & 20.10 \\
T3 & 15.19 & 26.13 & 27.50 \\
\hline
\multicolumn{4}{l}{\textit{Open-source models}} \\
ZS IndicConformer (Baseline-1) & 15.03 & 25.56 & 26.78 \\
FT IndicConformer (Baseline-2) & 11.40 & 19.01 & 20.23 \\
FT Whisper-large-v3 & 14.07 & 23.60 & 26.32 \\
\hline
\multicolumn{4}{l}{\textit{Closed-source models}} \\
Gemini 2.5 Pro & 10.76 & 19.60 & -- \\
Sarvam AI (Saaras v3) & 14.20 & 24.60 & 26.63 \\
\hline
\end{tabular}
}
% \vspace{-0.1in}
\end{table}

\subsection{Track 3: Topic Identification}
Table \ref{tab:challenge_results} presents the results for both Track $3$ and Track $4$. Track $3$ received two submissions in Eval Phase-I and both teams outperformed the baseline system described in Section \ref{ssec:ds-baseline}. 

\begin{table}[t]
\centering
\caption{Results for Track-3 (TI) and Track-4 (DS).}
% \vspace{-1em} 
\label{tab:challenge_results}
\resizebox{0.35\textwidth}{!}{
\begin{tabular}{lcc|lc}
\hline
\multicolumn{3}{c|}{\textbf{Track-3}} & \multicolumn{2}{c}{\textbf{Track-4}} \\
\hline
\textbf{Team}  & \textbf{R-1} & \textbf{R-L} & \textbf{Team} & \textbf{R-L} \\
\hline
T1 &  \textbf{0.46} & \textbf{0.44} & T1 & \textbf{0.20} \\
T2 &  0.36 & 0.35 & T2 & 0.19 \\
& & &   T3 & 0.16 \\
Baseline  & 0.15 & 0.14  & Baseline & 0.18 \\
Gemini 2.5 Pro  &  0.40  & 0.39 & Gemini 2.5 Pro & 0.21\\
\hline
\end{tabular}
}
\vspace{-0.05in}
\end{table}

Team 1 (T1) achieved the highest ROUGE scores, obtaining a ROUGE-1 of $0.46$ and a ROUGE-L of $0.44$, by using the Gemini 3 Pro end-to-end model operating directly on raw audio in a zero-shot configuration.
Team 2 (T2) enhanced the baseline ASR-LLM pipeline by improving translation, topic extraction and incorporating auxiliary patient information. They included supplementary cues from raw speech, such as vocal characteristics, to refine topic prediction; for instance, knowledge of a patient’s sex may help distinguish between stomach pain and menstrual pain.
They replaced the baseline translation model with Llama 3.1 70B Instruct and increased the decoding length to 2048 to better capture conversational context. For topic extraction, they employed Llama3.1-Aloe-Beta-8B, a model fine-tuned for medical reasoning with prompts refined iteratively and guided by a reference list of development-set topics. Additionally, diarization outputs were used to isolate patient speech, from which age and biological sex were estimated using a Wav2Vec2-based model. These attributes were appended to the prompt to enable age- and sex-aware topic identification. Their approach relies entirely on pre-trained models without additional fine-tuning, using only the released development sets for validation.

\subsection{Track 4: Dialogue Summarization}
Track $4$ received three submissions in Eval Phase-I, of which two teams outperformed the baseline system. The top-performing team (T1) proposed an end-to-end pipeline combining Silero-VAD \cite{Silero_VAD} for segmentation, speech translation from Hindi to English using AudioX-North Whisper \cite{jiviai_audiox_north_2024}, and GPT-4o-mini-based summarization using a structured clinical prompt. 
%Audio was segmented using Silero VAD, translated from Hindi to English using AudioX-North Whisper \cite{jiviai_audiox_north_2024} and summarized with GPT-4o-mini \cite{achiam2023gpt} using a structured clinical prompt. 
Team 2 (T2) used the Gemini-3 Pro end-to-end model in a $6$-shot configuration, where raw audio was directly provided as input without an intermediate ASR step and the model was guided by six example transcript–summary pairs from the challenge development set.

\section{Discussion and Conclusion} 
The DISPLACE-M challenge aims to study and investigate the difficulties that state-of-the-art systems face in transcribing and understanding frontline health conversations. Our results indicate that upstream tasks such as speaker diarization and automatic speech recognition still require improvements to enable seamless understanding of these conversations. The best ASR system achieved a WER of $18.2$ \% (and tcpWER of $18.65$ \%). Further, downstream tasks such as topic identification and dialogue summarization remain significantly more challenging due to their domain-specific nature and the complexity of interactions. 
Among the evaluated tasks, dialogue summarization is found to be the most challenging and we observe that even large closed-source models struggle to generate reliable and clinically accurate medical summaries. One key reason is the conversational structure of health worker-seeker interactions, as they often contain implicit symptoms, fragmented descriptions and context-dependent dialogue, which require deeper reasoning to interpret correctly. Generating accurate summaries requires domain knowledge and multi-step reasoning to interpret symptoms and identify medically relevant information. 
Since the timeline was only about $6$ weeks, we have initiated a Phase-II plan for this evaluation which allows $10$ more weeks of open-leader-board style evaluation. The Phase-II evaluation will include more languages (beyond Hindi).  

In summary, we designed a benchmark to advance conversational AI for frontline health dialogue understanding. 
The evaluation framework consists of four tracks - speaker diarization, automatic speech recognition, topic identification and dialogue summarization, allowing a comprehensive evaluation pipeline from speech processing to higher-level conversational understanding. We also described the data collection and annotation process, as well as baseline systems for the proposed tasks. Further, the paper summarizes the key findings from the Phase-I evaluation.

\bibliographystyle{IEEEtran}
\bibliography{mybib}

\appendix
\section{Appendix}

\subsection{DISPLACE-M Dataset Details}
\label{sec:dataset-details}

The primary goal of our data collection exercise was to build a diverse, high-quality, medical domain-specific, conversational speech dataset to support the development and evaluation of speech analytics solutions, such as automatic summarization, report generation, transcription, and information extraction.

\subsubsection{Participants and Setting}
\begin{figure}[!h]
\centering

\begin{subfigure}{0.5\textwidth}
    \centering
    \includegraphics[width=0.45\linewidth]{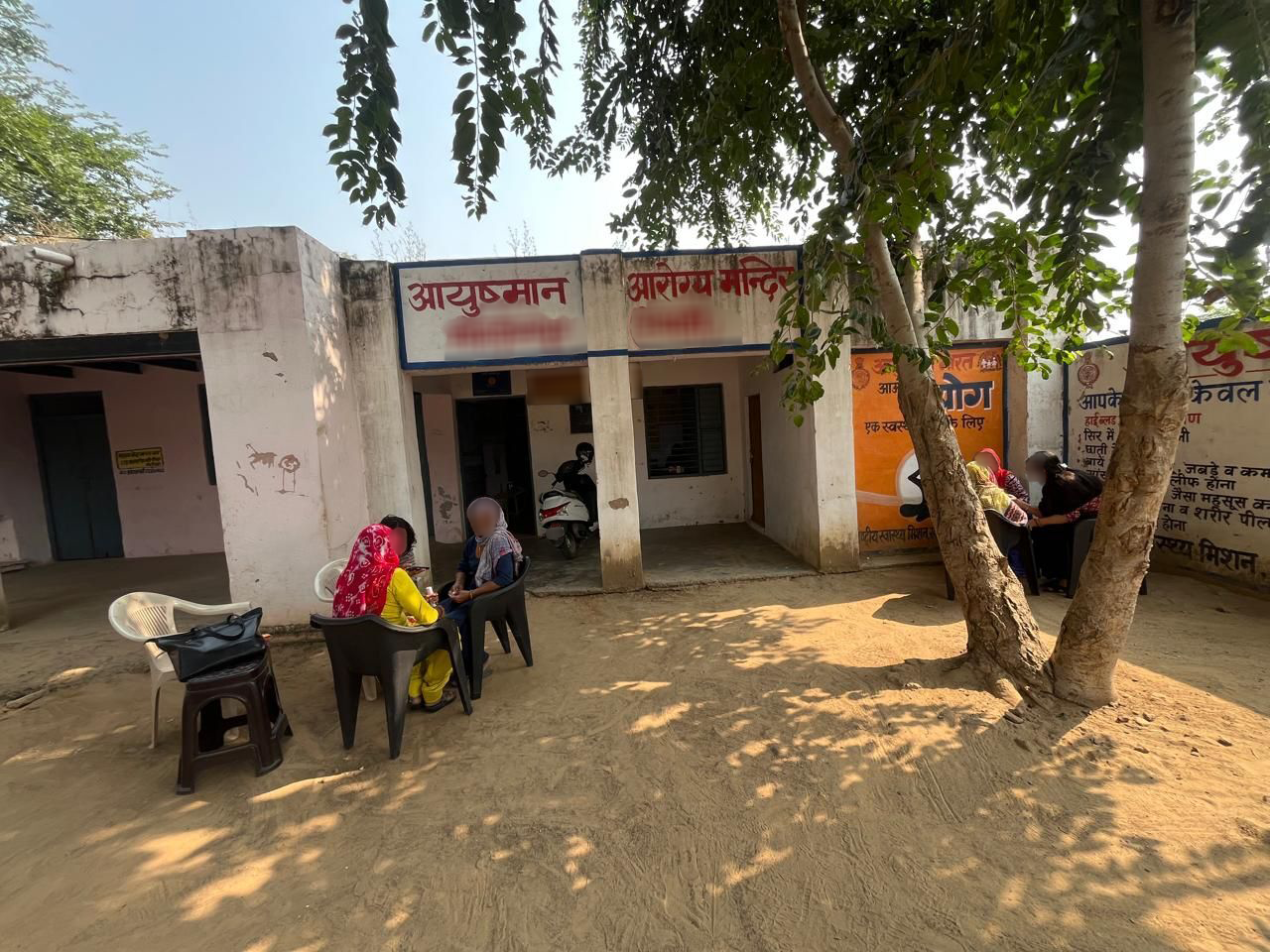}
    \includegraphics[width=0.45\linewidth]{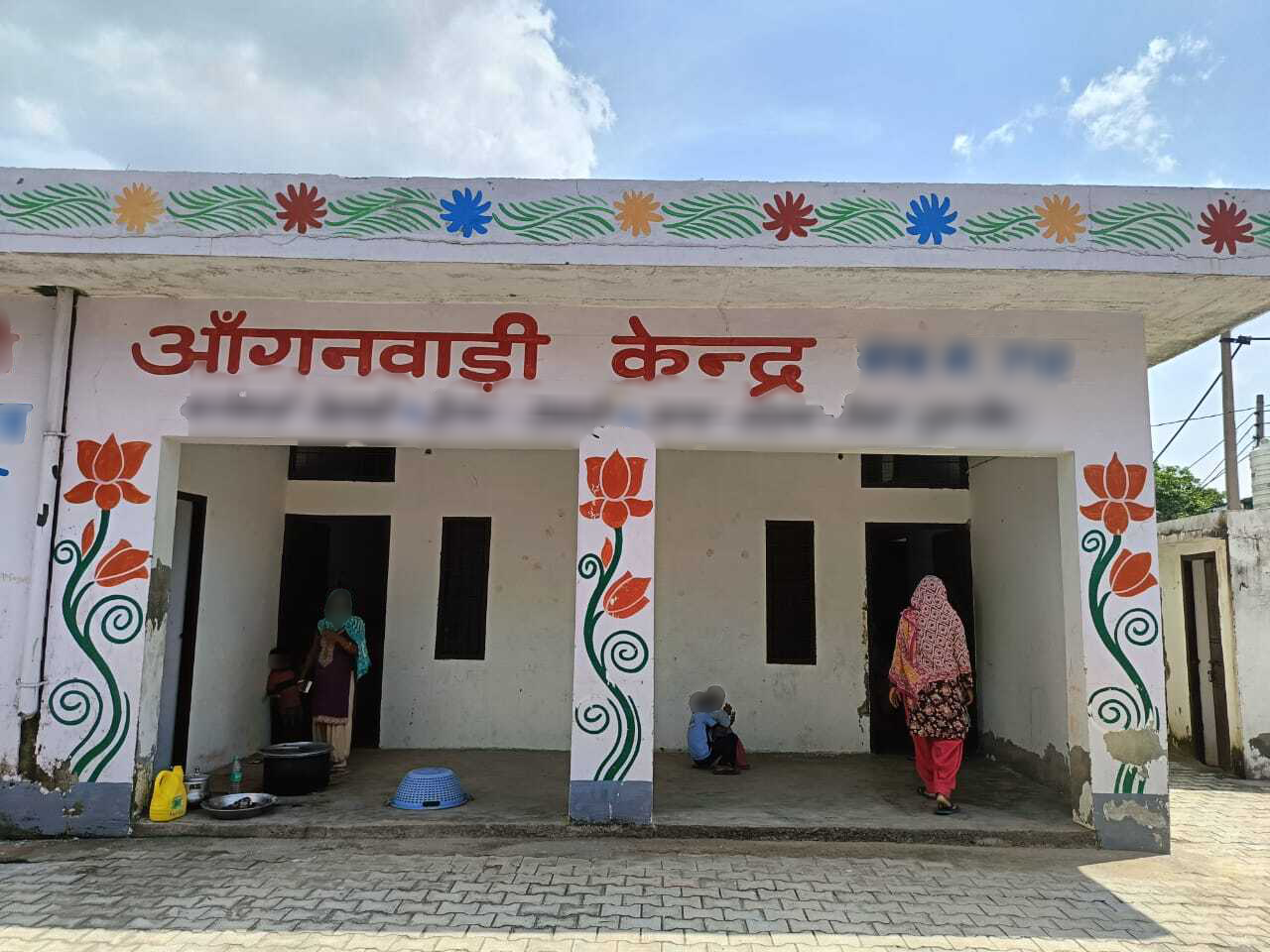}
    \includegraphics[width=0.45\linewidth]{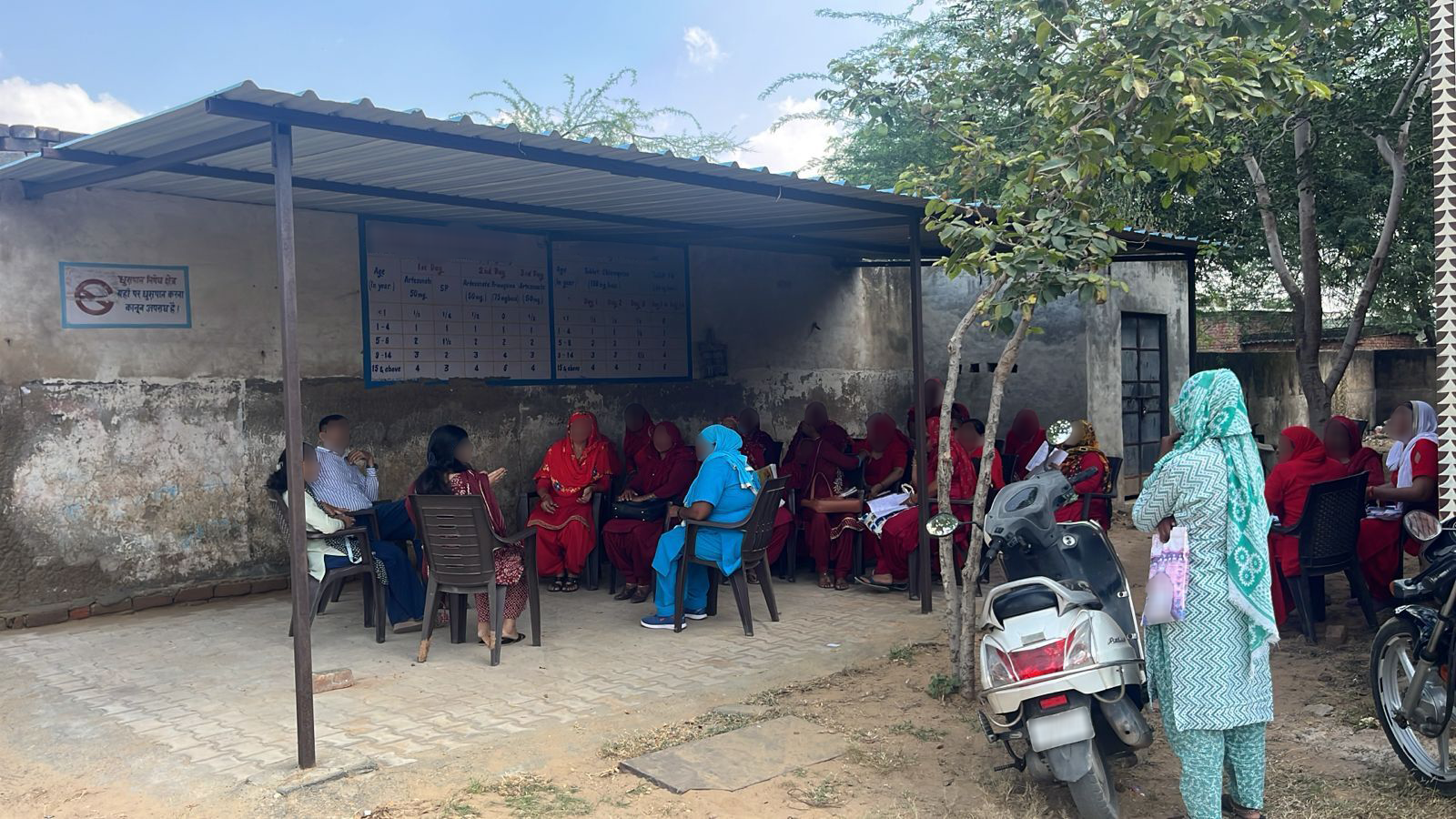}
    \includegraphics[width=0.45\linewidth]{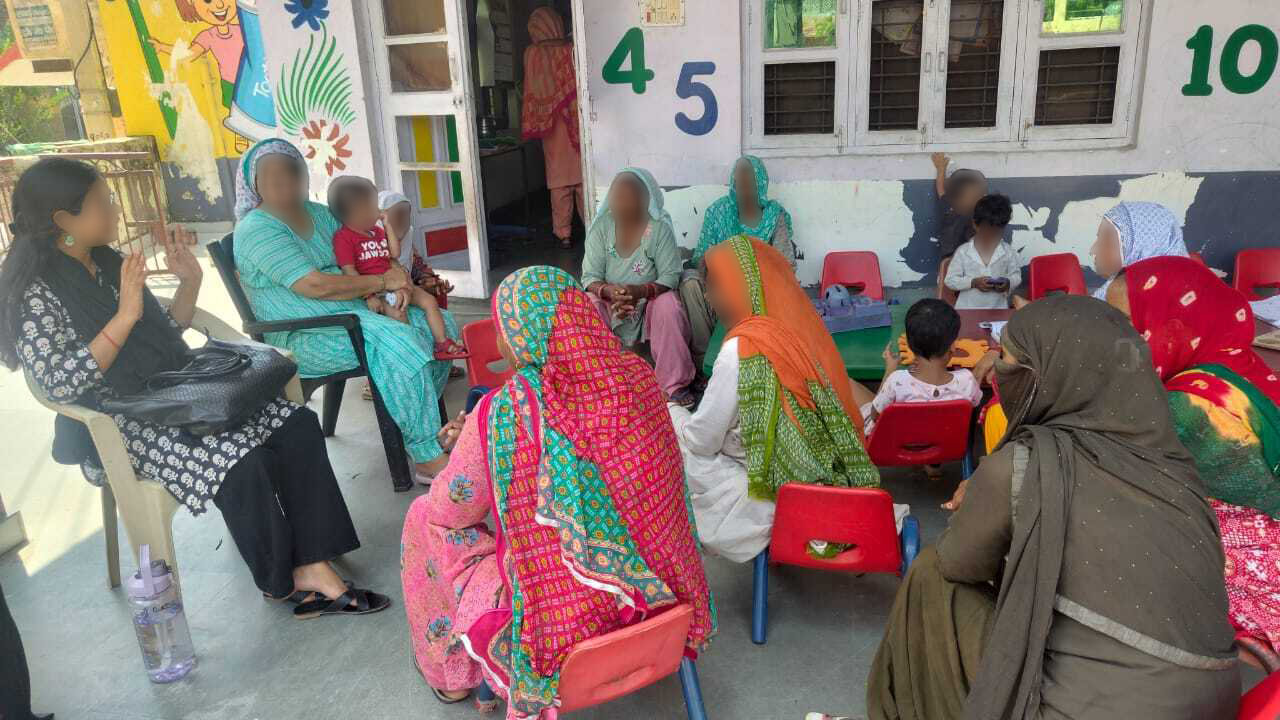}
    % \caption{Ayushman Arogya Kendra}
\end{subfigure}

\caption{Examples of dataset collection environments}
\label{fig:data_collection_env}
\end{figure}

The primary participants were ASHA workers and Anganwadi sevikas, two key cadres in India’s public health system. Recordings captured natural ASHA–health seekers conversations occurring during routine community healthcare interactions. These interactions took place in real care environments, including Anganwadi centers, primary schools, open village meeting spaces (chaupals), household visits, and community outreach settings as shown in Figure \ref{fig:data_collection_env}.

\begin{figure}[t]
\centering
\includegraphics[width=0.45\textwidth]{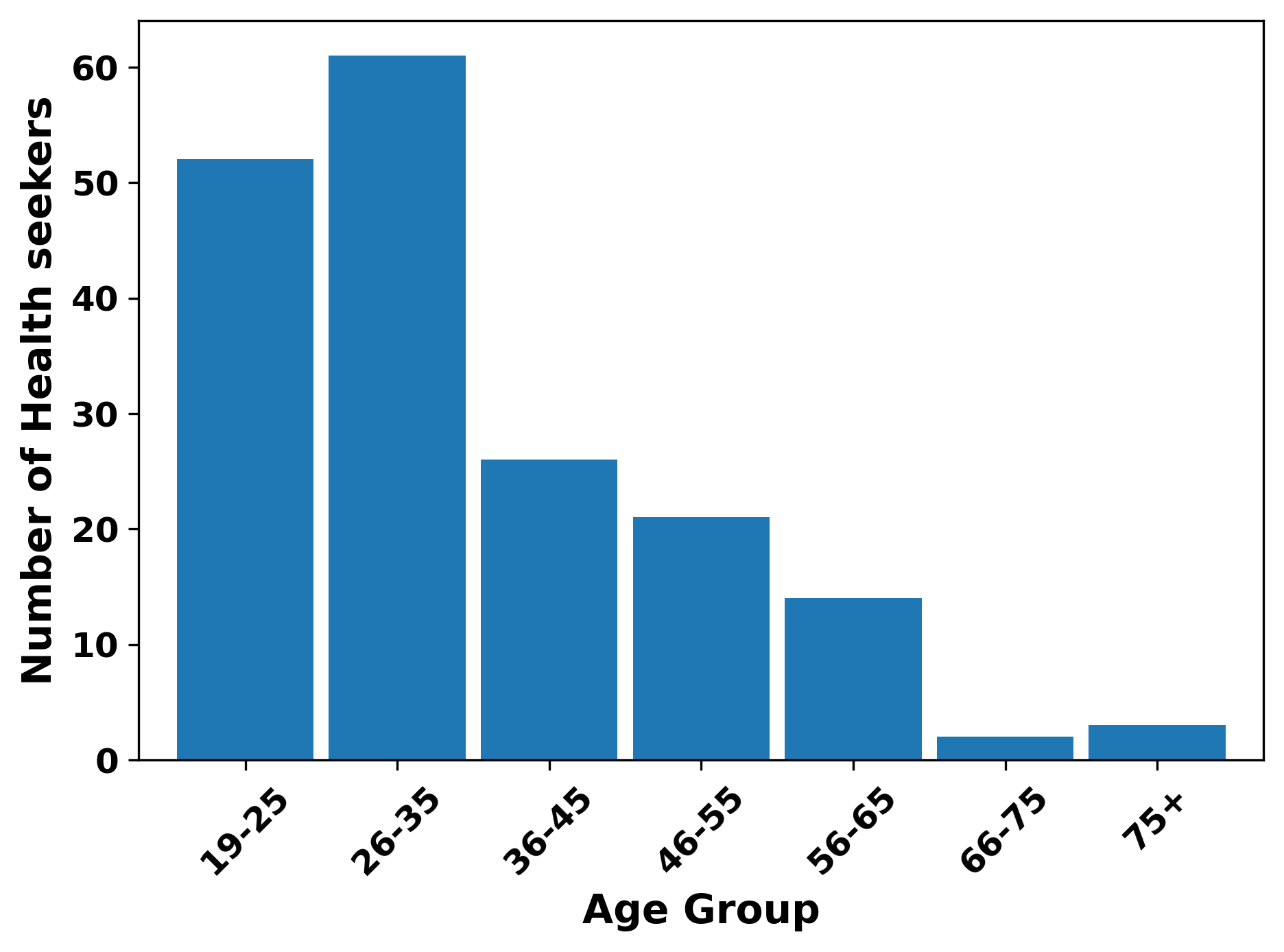}
\caption{Age distribution of health seeeker in the collected conversations.}
\label{fig:age_distribution}
\end{figure}

Figure~\ref{fig:age_distribution} illustrates the age distribution of health seekers represented in the collected conversations. The majority of health seekers fall within the age range of 19 to 35 years. The dataset mainly consists of female participants, accounting for 85\% of the health seekers, while male participants constitute 15\%.

\subsubsection{Conversational Domains and Linguistic Coverage}
\begin{figure*}[t]
  \centering
  \includegraphics[width=0.90\textwidth]{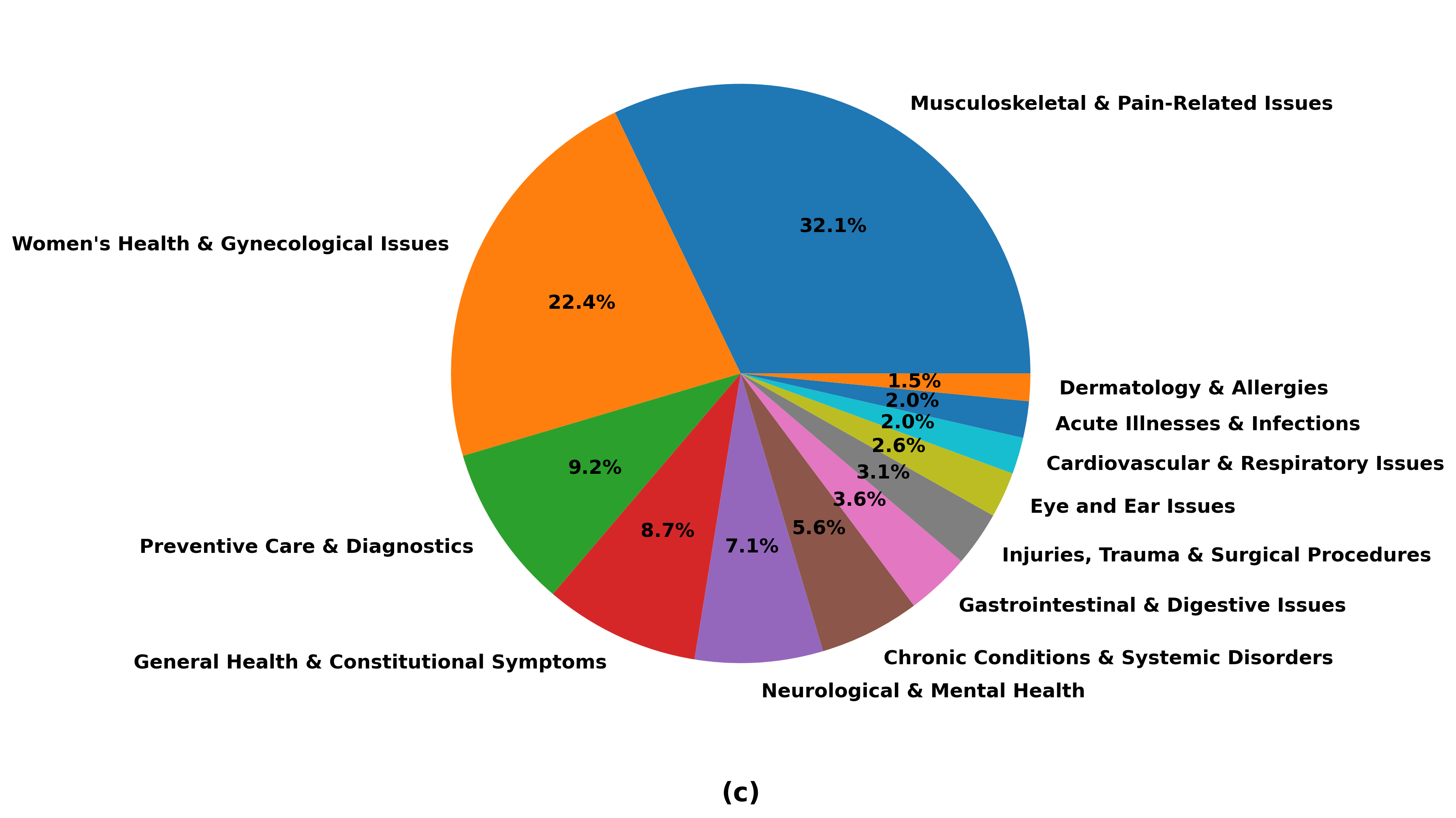}
  \caption{DISPLACE-M corpus statistics - Distribution of Topics}
  \label{fig:data_stats_topic}
\end{figure*}

Conversations reflected everyday healthcare concerns commonly addressed at the community level, including (as shown in Figure~\ref{fig:data_stats_topic}):
\begin{enumerate}
    \item Chronic musculoskeletal pain
    \item Maternal and menstrual health
    \item Fever and respiratory infections
    \item Metabolic conditions such as hypertension and diabetes
\end{enumerate}

\subsubsection{Metadata and Documentation}
Each recording was paired with a standardized on-site metadata schema to capture contextual and operational attributes. Metadata fields included:
\begin{itemize}
    \item Speaker role (health worker or health seeker)
    \item Basic demographic indicators
    \item Language and dialect information
    \item Geographic location
    \item Recording environment (indoor/outdoor, individual/group)
    \item Device characteristics
    \item Topical healthcare domain tags
\end{itemize}
This metadata enables downstream analysis of linguistic variation, acoustic conditions, and domain-specific performance.

\subsubsection{Field Onboarding and Protocol}
Prior to data collection, participating health workers underwent capacity-building and rehearsal-based onboarding. These one-on-one sessions covered 
\begin{itemize}
    \item Step-by-step recording protocols
    \item Informed consent and privacy safeguards
    \item Purpose and scope of data collection 
    \item Procedures for refusals and handling sensitive topics.
\end{itemize} 

The data collection followed a structured execution flow consisting of site preparation (staff briefing, environment readiness, and recording setup), on-ground onboarding, recording execution and post-recording validation to ensure ethical and procedural compliance.

\subsubsection{Operational Challenges and Mitigations}
Field deployment surfaced several practical challenges that shaped execution:
\begin{enumerate}
    \item Community trust and participation: Trust-building often required repeated outreach, including door-to-door engagement to address privacy concerns. Established ASHA–community relationships facilitated open dialogue, particularly with women, while engaging male participants typically required additional rapport-building.
    \item Social context and self-consciousness: Many interactions occurred in communal settings where participants were initially hesitant to speak openly. To preserve naturalness, teams minimized visible intervention and allowed ASHA workers to lead conversations, reducing recording-induced behavior shifts.
   \item Acoustic variability: Recordings spanned indoor centers, outdoor gatherings, and household visits, resulting in diverse acoustic conditions. Group settings occasionally introduced overlapping or multi-party speech, leading to turn-taking ambiguity and interruptions.
    \item Privacy and safety: Privacy-safe capture was a strict operational priority. Teams actively avoided recording personally identifiable information (e.g., names, phone numbers) and paused or redirected conversations when such content arose.
    \item Topic sensitivity: Discussions involving reproductive health or chronic conditions required careful pacing, reassurance, and discretion to maintain participant comfort and ethical integrity.
\end{enumerate}

% \begin{figure}[t]
%   \centering
%   \includegraphics[width=0.40\textwidth]{figures/age_distribution.png}
%   \caption{DISPLACE-M corpus statistics - Patient age distribution.}
%   \label{fig:data_stats_1}
% \end{figure}

\begin{figure*}
    \centering
    \includegraphics[width=\linewidth]{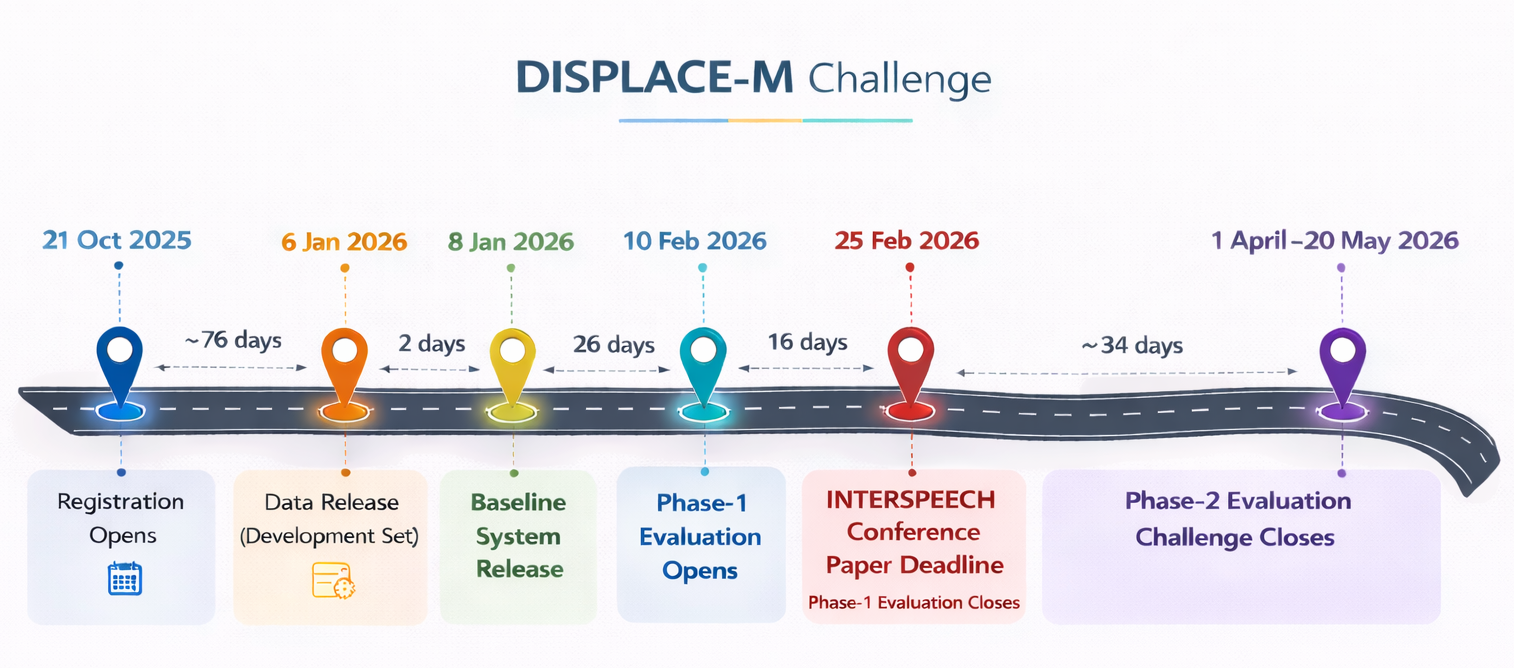}
    \caption{Timeline of DISPLACE-M}
    \label{fig:timeline}
\end{figure*}

\subsection{Annotation, Segmentation, and Quality Control Framework}
\label{ssec:app_data_annotation}
The annotation workflow:
\subsubsection{Stage 1: Manual Segmentation (Three-Step Protocol)}
\begin{enumerate}
    \item Step 1: Discourse-Based Boundary Identification
        \begin{itemize}
            \item Trained annotators manually identified natural conversational boundaries based on discourse structure, speaker turns, pauses, and interactional transitions.
            \item Segmentation was conversation-driven rather than fixed-duration based, preserving contextual and pragmatic continuity.
            \item Each segment was precisely time-aligned and enriched with structured metadata, including speaker identifiers, segment IDs, and start–end timestamps.
        \end{itemize}
    \item Step 2: Structural and Temporal Validation
    \begin{itemize}
        \item Independent review of segment boundaries to verify timestamp precision and prevent truncation or spillover across speaker turns.
        \item Cross-verification of speaker attribution and conversational sequencing to ensure structural coherence.
    \end{itemize}
    \item Step 3: Consistency and Alignment Audit
    \begin{itemize}
        \item Final validation pass assessing inter-annotator agreement, boundary consistency, and temporal integrity across the full recording.
        \item Discrepancies were resolved through adjudication to minimize segmentation variance and ensure standardization.
        \item Segments failing to meet validation criteria were returned for re-segmentation prior to further processing.
    \end{itemize}
    Only segments that satisfied all three validation layers were advanced to transcription.
\end{enumerate}

\subsubsection{Stage 2: Transcription (TR – Verbatim Rendering)}
\begin{itemize}
    \item Initial draft transcripts were generated using pre-existing ASR architectures that were fine-tuned by our team.
    \item These machine-generated outputs served as preliminary annotations 
    \item Expert human transcribers systematically reviewed and corrected the drafts to ensure strict verbatim fidelity.
    \item Transcription was performed in Hindi, Haryanvi, and Bhojpuri, faithfully reflecting dialectal forms, pronunciation-driven variations, and localized lexical usage.
    \item All transcripts were rendered in the Devanagari script to maintain orthographic standardization across linguistic varieties.
    \item The protocol preserved repetitions, disfluencies, pauses, overlaps, and interactional markers as present in the audio signal.
\end{itemize}

\subsubsection{Stage 3: Multi-Layer Quality Control (QC1–QC5)}
\begin{itemize}
    \item QC1 – Format and Completeness Review: 
    Verification of structural formatting, transcript completeness, normalization consistency, and preliminary PII screening
    \item QC2 – Consistency and Risk-Based Audit: 
    Rubric-based spot reviews combined with stratified sampling informed by risk indicators (e.g., acoustic variability, multi-party speech, overlap density).
    \item QC3 – Sound Tag Validation
    Verification and correction of non-speech markers and contextual acoustic annotations.
    \item QC4 – Overlap and Pause Verification
    Detailed examination of overlapping speech, interruptions, and pause representation to ensure accurate modeling of conversational dynamics.
    \item QC5 – Final Compliance and Release Check
    End-stage validation ensuring alignment between audio, timestamps, and speaker labels, along with final ethical and compliance screening prior to dataset release.
\end{itemize}
This methodology balances field realism with structured execution, enabling the collection of conversational speech that reflects authentic healthcare interactions in low-resource settings while maintaining ethical rigor and documentation consistency.

%Professional annotators were hired to label the data verbatim. 
% . Annotations include speaker diarization, language tags (with allowance for code-mixing), and markers for overlapping speech or background noise. Each audio recording is accompanied by a metadata entry stored in an .xlsx file. The metadata includes anonymized speaker IDs, language and dialect information, demographic information, recording environment (indoor/outdoor, individual/group) and topical healthcare domain tags.

\subsection{DISPLACE-M Timeline}
Figure \ref{fig:timeline} illustrates the timeline of the DISPLACE-M Challenge, outlining the key milestones - registration phase, dataset release, system development period and evaluation phases.

\subsection{Evaluation Metric}
Each track in the DISPLACE-M Challenge was evaluated using task-specific metrics. Speaker diarization performance was measured using Diarization Error Rate (DER), while automatic speech recognition was evaluated using Time-Constrained Minimum-Permutation Word Error Rate (tcpWER). Downstream tasks - topic identification and dialogue summarization were evaluated using ROUGE-based metrics.

\subsubsection{Diarization Error Rate (DER)}
\label{subsec:DER}

DER is the main metric used to evaluate speaker diarization systems in the literature. As described in the NIST Rich Transcription Spring 2003 Evaluation (RT03S) \footnote[1]{\url{https://web.archive.org/web/20160805233512/http://www.itl.nist.gov/iad/mig/tests/rt/2003-spring/index.html}}, DER is defined as follows:

\begin{equation}
    DER = \frac{D_{FA} + D_{miss} + D_{error}}{D_{total}}
\end{equation}

where, 

\begin{itemize}
    \item $D_{FA}$ represents the total system speaker duration which is not attributed to a reference speaker. 
\item $D_{miss}$ denotes the total reference speaker duration, which is not attributed to a system speaker. 
\item $D_{error}$ represents the total system speaker duration attributed to the wrong reference speaker. 
\item $D_{total}$ denotes the total reference speaker duration, which is represented as the summation of all the reference speakers segments' duration.
\end{itemize}

The DER metric will be calculated {\textbf{with overlap}} and {\textbf{without collar}}. Here, DER with overlap means that the segments containing the speech of multiple simultaneous speakers are included for evaluation. Also, DER without collar signifies that no tolerance around the actual speaker boundaries is considered during assessment. 

The speaker diarization system evaluation will be done using dscore (version 1.0.1) script, which is available at:{\url{https://github.com/nryant/dscore}}. 
Each participating team must upload a system output RTTM corresponding to each of the conversations present in the blind evaluation set. 
% The following nomenclature should be followed to score a bunch of system output RTTMs against their corresponding reference RTTMs:\\

\subsubsection{Time-Constrained Minimum-Permutation Word Error Rate (tcpWER)}
\label{subsec:tcpwer}
Time-Constrained Minimum-Permutation Word Error Rate (tcpWER)~\cite{neumann23_chime} is an evaluation metric used to assess the performance of multi-speaker automatic speech recognition (ASR) systems. The metric extends the conventional Word Error Rate (WER) by incorporating both temporal constraints and speaker permutation invariance, making it suitable for conversational speech involving multiple speakers.

In multi-speaker ASR scenarios, systems often produce speaker-attributed transcriptions where predicted speaker labels do not directly correspond to the reference speaker identities. A naive computation of WER would therefore unfairly penalize systems for speaker-label mismatches. tcpWER addresses this issue by searching for the optimal permutation of speaker assignments that minimizes the overall word error rate while preserving the temporal alignment of speaker turns.

The time-constrained formulation ensures that hypothesis words are considered correct only if they occur within an appropriate temporal window corresponding to the reference speaker segments. As a result, tcpWER reflects not only transcription accuracy but also the impact of speaker attribution and segmentation errors. Consequently, the metric captures the combined effect of speaker diarization quality and ASR performance on the final transcription output.

In the DISPLACE-M Challenge, ASR systems are evaluated using the tcpWER implementation provided by the \textit{meeteval} toolkit\footnote{\url{https://github.com/fgnt/meeteval}}. Participating teams are required to submit transcription output files corresponding to each recording in the blind evaluation set for scoring.

\subsubsection{ROUGE-1}
\label{subsec:rouge-1}

ROUGE-1~\cite{lin-2004-rouge} is an automatic evaluation metric used to measure the overlap of unigrams (individual words) between a system-generated text and a reference text. It is one of the simplest and most widely adopted variants of the ROUGE family and provides an indication of how much content from the reference is captured by the generated output.

Given a reference text and a hypothesis text, ROUGE-1 computes the number of matching unigrams between the two. Based on the unigram overlap, precision and recall are defined as follows:

\begin{equation}
    P_{ROUGE-1} = \frac{\text{Number of overlapping unigrams}}{\text{Total unigrams in hypothesis}}
\end{equation}

\begin{equation}
    R_{ROUGE-1} = \frac{\text{Number of overlapping unigrams}}{\text{Total unigrams in reference}}
\end{equation}

The ROUGE-1 F-score is then calculated as the harmonic mean of precision and recall:

\begin{equation}
    ROUGE\text{-}1 = \frac{2 \cdot P_{ROUGE-1} \cdot R_{ROUGE-1}}{P_{ROUGE-1} + R_{ROUGE-1}}
\end{equation}

ROUGE-1 primarily captures lexical overlap between the system output and the reference text without considering word order or sequence structure. As a result, it is effective for measuring content coverage and identifying whether key terms present in the reference are also produced by the system.

\subsubsection{Rouge-L}
Rouge-L~\cite{lin-2004-rouge} is a standard metric widely used in the literature for the automatic evaluation of summarization tasks. It measures the similarity between the generated (hypothesis) summary and a reference summary based on Longest Common Subsequence (LCS) by identifying the longest sequence of tokens which appears in both summaries in the same order not necessarily contiguously, capturing the sentence level structural similarity while remaining tolerant to minor variations in wording. \par
Let $X = (x_1, x_2, \dots, x_m)$ be the reference summary and $Y = (y_1, y_2, \dots, y_n)$ be the generated (hypothesis) summary. Then, the Rouge-L is calculated as defined in \autoref{rouge-l}.

\begin{equation}
    \label{rouge-l}
    \mathrm{ROUGE\text{-}L} = \frac{(1 + \beta^2)\, P_{\mathrm{LCS}}\, R_{\mathrm{LCS}}}{R_{\mathrm{LCS}} + \beta^2 P_{\mathrm{LCS}}}
\end{equation}
Here, $P_{\mathrm{LCS}}$ and $R_{\mathrm{LCS}}$ are defined as:
\begin{equation}
P_{\mathrm{LCS}} = \frac{\mathrm{LCS}(X,Y)}{|Y|}, \quad 
R_{\mathrm{LCS}} = \frac{\mathrm{LCS}(X,Y)}{|X|},
\end{equation}
where $|X|$ and $|Y|$ are the lengths of the reference summary $X$ and the generated summary $Y$, respectively. The parameter $\beta$ controls the relative importance of recall versus precision and, in practice, to give equal weight to precision and recall, $\beta$ is usually set to 1.

In the DISPLACE-M challenge, Dialogue Summarization systems are evaluated using the Rouge-L implementation provided by the \texttt{rouge\_score}\footnote{\url{https://pypi.org/project/rouge-score/}} library. Participating teams are required to submit summarization output files corresponding to each recording in the blind evaluation set for scoring.

\subsection{Prompts Used for LLM-based Systems}
This section provides the prompts used for the LLM-based components in the baseline systems - topic identification and dialogue summarization.

\subsubsection{Task 3: Topic Identification}
\label{ssec:task3_topic_id}

we employ a constrained zero-shot prompting strategy.The prompt template is defined as follows:

\begin{quote}
\small
\texttt{You are a medical classifier.\\
Task: Extract only the health seeker's ongoing health problems.\\
STRICT INSTRUCTIONS:\\
1. Exclude family history and resolved past illnesses.\\
2. Do not include advice, causes, or explanations.\\
3. Return a comma-separated list of concise medical terms.\\
4. If no ongoing health problems are identified, return: "None".\\
Conversation: \{ASR-Text\}\\
Health Problems:
}
\end{quote}

\subsubsection{Track 4: Dialogue Summarization}
\label{ssec:task4_ds}
We use the following system instruction for the dialogue summarization task:
\begin{verbatim}
You are a Summary Generator.
\end{verbatim}
We use the following user prompt template for the dialogue summarization task:
\begin{lstlisting}
{ASR-Text}

Task: Summarize the above conversation in English as a concise, factual patient summary and follow the format with these two tags <SUMMARY> and </SUMMARY>, the summary should be between these two tags and there should not be anything extra. 

STRICT INSTRUCTIONS:
1. Use third-person narration only (e.g., "The patient reports...", "The doctor advises...").
2. Include ONLY information explicitly mentioned in the conversation.
3. Structure the summary to cover: Chief Complaint, History of Present Illness, Diagnosis (if any), and Treatment Plan.
4. Include objective measurements (e.g., BP, temperature) and medications if stated.
5. Do NOT hallucinate or add outside information.
6. The output must be concise and clinically accurate.
\end{lstlisting}
Here, \verb|{ASR-Text}| is a placeholder for the ASR transcript of the conversation obtained from the ASR-Baseline system.

\end{document}